\RequirePackage{fix-cm}
\documentclass[smallextended]{svjour3}       
\smartqed  
\usepackage[ruled,lined]{algorithm2e}
\usepackage{multirow}
\usepackage{xfrac}
\usepackage{subfigure}
\usepackage{epstopdf}
\usepackage{graphicx}
\usepackage{amsfonts}
\usepackage{amsmath}
\usepackage{amssymb}
\begin{document}

\title{Incorporating Semantic Knowledge into \\Latent Matching Model in Search}

\catcode`\<=\active \def<{
\fontencoding{T1}\selectfont\symbol{60}\fontencoding{\encodingdefault}}
\catcode`\>=\active \def>{
\fontencoding{T1}\selectfont\symbol{62}\fontencoding{\encodingdefault}}
\newcommand{\nocomma}{}
\newcommand{\noplus}{}
\newcommand{\tmop}[1]{\ensuremath{\operatorname{#1}}}

\newcommand{\tmfloatcontents}{}
\newlength{\tmfloatwidth}
\newcommand{\tmfloat}[5]{
  \renewcommand{\tmfloatcontents}{#4}
  \setlength{\tmfloatwidth}{\widthof{\tmfloatcontents}+1in}
  \ifthenelse{\equal{#2}{small}}
    {\ifthenelse{\lengthtest{\tmfloatwidth > \linewidth}}
      {\setlength{\tmfloatwidth}{\linewidth}}{}}
    {\setlength{\tmfloatwidth}{\linewidth}}
  \begin{minipage}[#1]{\tmfloatwidth}
    \begin{center}
      \tmfloatcontents
      \captionof{#3}{#5}
    \end{center}
  \end{minipage}}

\author{ Shuxin Wang \and Xin Jiang \and Hang Li \and Jun Xu \and Bin Wang}

\institute{Shuxin Wang \at Institute of Computing Technology,  Chinese Academy of Sciences\\
\email{wangshuxin@ict.ac.cn}
\and
 Xin Jiang \at Huawei Noah's Ark Lab\\
\email{jiang.xin@huawei.com}
\and
Hang Li \at Huawei Noah's Ark Lab\\
\email{hangli.hl@huawei.com}
\and
Jun Xu \at Institute of Computing Technology,  Chinese Academy of Sciences\\
\email{junxu@ict.ac.cn}
\and
Bin Wang \at Institute of Information Engineering, Chinese Academy of Sciences\\
\email{wangbin@iie.ac.cn}
}

\maketitle
\begin{abstract}
The relevance between a query and a document in search can be represented as matching degree between the two objects. Latent space models have been proven to be effective for the task, which are often trained with click-through data. One technical challenge with the approach is that it is hard to train a model for tail queries and tail documents for which there are not enough clicks. In this paper, we propose to address the challenge by learning a latent matching model, using not only click-through data but also semantic knowledge. The semantic knowledge can be categories of queries and documents as well as synonyms of words, manually or automatically created.  Specifically, we incorporate semantic knowledge into the objective function by including regularization terms.  We develop two methods to solve the learning task on the basis of coordinate descent and gradient descent respectively, which can be employed in different settings. Experimental results on two datasets from an app search engine demonstrate that our model can make effective use of semantic knowledge, and thus can significantly enhance the accuracies of latent matching models, particularly for tail queries.
\end{abstract}

\keywords{Latent Matching Model, Semantic Knowledge, Learning to Match, Regularized Mapping to Latent Structures}

\section{Introduction}\label{sec:introduction}
In search, given a query documents are retrieved and ranked according to their relevance, which can be represented by the matching score between the query and each of the documents, referred to as semantic matching in~\cite{LiXu}. Traditional IR models, including Vector Space Model (VSM)~\cite{Salton:1975is}, BM25~\cite{robertson1995okapi}, and Language Models for Information Retrieval (LMIR)~\cite{ponte1998language,zhai2004study} can be viewed as matching models for search, created without using machine learning. The models work well to some extent, but they sometimes suffer from mismatch between queries and documents.

Recently significant effort has been made on automatic construction of matching models in search, using machine learning and click-through data. The learned models can effectively deal with mismatch and outperform traditional IR models~\cite{LiXu}. Among the proposed approaches, learning a latent space model for matching in search becomes the state-of-the-art. The class of semantic matching models, called latent matching models in this paper, map the queries and documents from their original spaces to a lower dimensional latent space, in which  the matching scores are calculated as inner products of the mapped vectors.

Despite the empirical success of latent matching models, the problem of query document mismatch in search is still not completely solved. Specifically, it remains hard to effectively train a matching model which works well not only for frequent queries and documents, but also for rare queries and documents, because there is not sufficient click data for rare queries and documents. This in fact belongs to the long tail phenomenon, which also exists in many different tasks in web search and data mining.

One way to conquer the challenge would be to incorporate additional semantic knowledge into the latent matching models. Specifically, semantic knowledge about synonyms and categories of queries and documents can make the latent space better represent similarity between queries and documents. Suppose that ``Sacramento'' and ``capital of California'' are synonyms and it would be difficult to observe their association directly from click information
(e.g., a query and the title of clicked document), because both rarely occur in the data. If we can embed the knowledge into the learned latent space, then it will help to make judgment on the matching degrees between queries and documents containing the synonyms. The technical question which we want to address in this paper is how to incorporate semantic knowledge in the learning of latent space model in a theoretically sound and empirically effective way.

In this paper, as the first step of the work, we propose a novel method for learning a linear latent matching model for search, leveraging not only click-through data, but also semantics knowledge such as synonym dictionary and semantic categories. The semantic knowledge can either be automatically mined or manually created. Specifically, we reformulate the learning of latent space model by adding regularization, in which way semantic knowledge can be naturally embedded into the latent space and be utilized in matching. The learning problem becomes maximization of the matching degrees of relevant query document pairs as well as the agreement with the given semantic knowledge. Regularization is also imposed on the linear mapping matrices as well as their product in order to increase the generalization ability of the learned model.

For semantic knowledge acquisition, we employ two methods. In the first method, synonyms are automatically mined from click-through data. In the second method, the documents are assigned semantic categories in a hierarchy. In both methods, the semantic knowledge can be easily added into the proposed framework.

Without loss of generality, we take Regularized Mapping in Latent Space (RMLS)~\cite{Wu:2013wm}, one of the state-of-the-art methods for query document matching, as the basic latent matching model and augment it with semantic knowledge. We improve the optimization procedure of RMLS by introducing a new regularization term.
We further develop a coordinate descent algorithm and a gradient descent algorithm to solve the optimization problem. The algorithms can be employed in different settings and thus the learning can be generally carried out in an efficient and scalable way.

We conduct experiments on two large-scale datasets from a mobile app search engine. The experimental results demonstrate that our model can make effective use of the semantic knowledge, and significantly outperform existing matching models.

The contributions of the work include (1) proposal of a method for incorporating semantic knowledge into latent matching model, (2) proposal of two optimization methods to perform the learning task, (3) empirical verification of the effectiveness of the method, and (4) improvement of existing method of RMLS. We take the work as the first step toward the goal of combining machine learning and prior knowledge in learning of matching model.

The rest of the paper is organized as follows. After an introduction to related work in Section \ref{sec:related-work}, we describe the formulation of our latent matching model and the corresponding algorithm in Section \ref{sec:lmm-ir}. Section \ref{sec:lmm-knowledge} presents the latent matching model for incorporating semantic knowledge as well as our methods to acquire the knowledge. Experimental results and discussions are given in Section \ref{sec:experiments}. Section \ref{sec:conclusion} concludes this paper and gives future work.

\section{Related Work}\label{sec:related-work}
Matching between queries and documents is of central importance to search~\cite{LiXu}. Traditional information retrieval models including Vector Space Model (VSM)~\cite{Salton:1975is}, BM25~\cite{robertson1995okapi}, and Language Models for Information Retrieval (LMIR)~\cite{ponte1998language,zhai2004study} are based on term matching and may suffer from term mismatch.

Topic modeling techniques aim to discover the topics as well as the topic representations of documents in the document collection, and can be used to deal with query document mismatch. Latent semantic indexing (LSI)~\cite{Deerwester:1990via} is one typical non-probabilistic topic model, which decomposes the term-document matrix by Singular Value Composition (SVD) under the assumption that topic vectors are orthogonal. Regularized Latent Semantic Indexing (RLSI)~\cite{Wang:2013ux} formalizes topic modeling as matrix factorization with regularization of $\ell_1$/$\ell_2$-norm on topic vectors and document representation vectors. Probabilistic Latent Semantic Indexing (PLSI)~\cite{hofmann1999probabilistic} and Latent Dirichlet Allocation (LDA)~\cite{Blei:2003tn} are two widely used probabilistic topic models, in which each topic is defined as a probability distribution over terms and each document is defined as a probability distribution over topics. By employing one of the topic models, one can project queries and documents into the topic space and calculate their similarities in the space. However, topic modeling does not directly learn the query document matching relation, and thus its ability of dealing with query document mismatch is limited.

Several latent matching models are proposed to address the issue. In a latent matching model, queries and documents are deemed as objects in two different spaces and are mapped into the same latent space for matching degree calculation (e.g., inner product). The learning of the mapping functions is performed by using training data such as click-through log in a supervised fashion, and thus is more effective to deal with mismatch. Partial Least Square (PLS)~\cite{Rosipal:2005vw} is a method developed in statistics and can be utilized to model the matching relations between queries and documents. PLS is formalized as learning of two linear projection functions represented by orthonormal matrices and can be solved by Singular Value Decomposition (SVD). Canonical Correspondence Analysis (CCA)~\cite{hardoon2004canonical} is an alternative method to PLS. The difference between CCA and PLS is that CCA takes cosine as the similarity measure and PLS takes inner product as the similarity measure. Bai et al.~\cite{bai2009supervised} propose Supervised Semantic Indexing(SSI), which makes use of a pairwise loss function and learns a low-rank model for matching and ranking. Wu et al.~\cite{Wu:2013wm} propose a general framework for learning to match heterogeneous objects, and a matching model called Regularized Mapping to Latent Structures (RMLS) is specified. RMLS extends PLS by replacing its orthonormal constraints with $\ell_1$ and $\ell_2$ regularization. RMLS is superior to PLS in terms of computation efficiency and scalability.

Recently, non-linear matching models have also been studied. For example, Huang et al.~\cite{huang2013learning} propose a model referred to as Deep Structured Semantic Model (DSSM), which performs semantic matching with deep learning techniques. Specifically, the model maps the input term vectors into output vectors of lower dimensions through a multi-layer neural network, and takes cosine similarities between the output vectors as the matching scores. Lu and Li~\cite{lu2013deep} propose a deep architecture for matching short texts, which can also be queries and documents. Their method learns matching relations between words in the two short texts as a hierarchy of topics and takes the topic hierarchy as a deep neural network.

As another widely studied area, collaborative filtering (CF) is essentially user and item matching, which shares much similarity with query document matching in search. Latent factor models are state-of-the-art methods for collaborative filtering~\cite{abernethy2009new,agarwal2009regression,hofmann2004latent}. They are basically models for performing matching between users and items in a latent space. To enhance the accuracy of collaborative filtering, Ma et al.~\cite{Ma:2013ae} propose incorporating implicit social information (e.g., association between users) into the latent factor models. We will show the difference between our model and the latent factor models in CF in the next section.

Matching between query and document can be generalized as a more general machine learning task, which is indicated in \cite{LiXu} and referred to as learning to match. Learning to match aims to study and develop general machine learning techniques for various applications such as search, collaborative filtering, paraphrasing and textual entailment, question answering, short text conversation, online advertisement, and link prediction, and to improve the state-of-the-art for all the applications. The method proposed in this paper is potentially applicable to other applications as well.

\section{Latent Matching Model}\label{sec:lmm-ir}

\subsection{Problem Formulation}\label{sec:problem-formulation}
Let $\mathcal{X} \subset \mathbb{R}^{d_x}$ and $\mathcal{Y} \subset \mathbb{R}^{d_y}$ denote the two spaces for matching, and $x \in \mathcal{X}$ and $y \in \mathcal{Y}$ denote the objects in the spaces. In search, $x$ and $y$ are a query vector and a document vector, respectively. Suppose that there is a latent space $\mathcal{L} \subset \mathbb{R}^d$. We intend to find two mapping functions that can map the objects in both $\mathcal{X}$ and $\mathcal{Y}$ into $\mathcal{L}$.
When the two mapping functions are linear, they can be represented as matrices: $L_x \in \mathbb{R}^{d \times d_x}$ and $L_y \in \mathbb{R}^{d \times d_y}$. The degree of matching between objects $x$ and $y$ is then defined as inner product of $L_x x$ and $L_y y$:
\begin{equation}\label{eq:matching}
\mathrm{match}(x,y) = \langle L_x x, L_y y \rangle = x^T L_x^T L_y y.
\end{equation}
To learn the linear mappings, we need training data that indicates the matching relations between the objects from the two spaces.
In search, click-through logs are often used as training data, because they provide information about matching between queries and documents.
Following the framework by Wu et al.~\cite{Wu:2013wm}, given a training dataset of positive matching pairs $\{(x_i,y_i)\}_{i=1}^n$, the learning problem is formalized as
\begin{equation}\label{eq:matching-framework}
\begin{split}
\arg \max_{L_x, L_y}  \frac{1}{n}\sum_{i=1}^{n} {x_i}^T L_x^T L_y y_{i}, \\
\mathrm{subject\,to\,} L_x \in \mathcal{H}_x, L_y \in \mathcal{H}_y.
\end{split}
\end{equation}
where $\mathcal{H}_x$ and $\mathcal{H}_y$ denote the hypothesis spaces for the linear mappings $L_x$ and $L_y$, respectively. This framework subsumes Partial Least Square (PLS) and Regularized Mapping to Latent Structure (RMLS) as special cases.
For PLS, the hypothesis spaces are confined to matrices with orthonormal rows, i.e.,
\[
\begin{split}
\mathcal{H}_x=\{L_x\,|\,L_x L_x^T = I^{d \times d}\},\\
\mathcal{H}_y=\{L_y\,|\,L_y L_y^T = I^{d \times d}\},
\end{split}
\]
where $I$ is the identity matrix. RMLS replaces the orthonormal assumption with sparsity constraints on $L_x$ and $L_y$. More specifically, the hypothesis spaces in RMLS become:
\[
\begin{split}
\mathcal{H}_x=\{L_x\,|\,\|l_{xu}\|_p\leqslant \tau_{x,p}, p=1,2, u=1,\ldots,d_x\},\\
\mathcal{H}_y=\{L_y\,|\, \|l_{yv}\|_p\leqslant \tau_{y,p}, p=1,2, v=1,\ldots,d_y\},
\end{split}
\]
where $l_{xu}$ is the $u$-th column vector of $L_x$ and $l_{yv}$ is the $v$-th column vector of $L_y$. The column vectors are actually latent representations of the elements in the original spaces, for instance, the terms in queries and documents.
$\|\cdot\|_p$ denotes $\ell_p$ norm, and both $\ell_1$ and $\ell_2$ are used in RMLS. $\tau_{x,p}$ and $\tau_{y,p}$ are thresholds on the norms.

We point out that RMLS is not very robust, both theoretically and empirically. Wu et al.~\cite{Wu:2013wm} prove that RMLS gives a degenerate solution with $\ell_1$ regularization only. Specifically, the solution of $L_x$ and $L_y$ will be matrices of rank one and all the column vectors $l_{xu}$ and $l_{yv}$ will be proportional to each other. Wu et al.~\cite{Wu:2013wm} propose addressing the problem with further $\ell_2$ regularization on $l_{xu}$ and $l_{yv}$. However, this does not solve the problem, which we will explain later in Section~\ref{sec:learning-algorithm}. Our experiments also show that RMLS tends to create degenerate solutions.

We notice that RMLS does not penalize the case in which any $x$ in one space matches any $y$ in the other space, which may happen even when $L_x$ and $L_y$ are sparse. To cope with the problem, we introduce additional constraints on the matching matrix $L_x^T L_y$, whose $(u,v)$-th element corresponds to the matching score between the $u$-th basis vector from $\mathcal{X}$ and the $v$-th basis vector from $\mathcal{Y}$.
Specifically, we add $\ell_1$ and $\ell_2$ norms on $L_x^T L_y$ as follows, which can limit the overall degree of matching any two objects.
\[
\|L_x^T L_y\|_1=\sum_{u,v} |l_{xu}^T l_{yv}| \textrm{, } \|L_x^T L_y\|_2^2=\sum_{u,v} (l_{xu}^T l_{yv})^2.
\]
This regularization can prevent the model from becoming a degenerate solution, and thus make the model more robust. With all of the constraints the hypothesis spaces of $L_x$ and $L_y$ become:
\[
\begin{split}
\mathcal{H}_x=\{L_x\,|\,\|l_{xu}\|_p\leqslant \tau_{x,p}, \|l_{xu}^T l_{yv}\|_p \leqslant \sigma_p, p=1,2, \forall u, v\},\\
\mathcal{H}_y=\{L_y\,|\,\|l_{yv}\|_p\leqslant \tau_{y,p}, \|l_{xu}^T l_{yv}\|_p \leqslant \sigma_p, p=1,2, \forall u, v\}.
\end{split}
\]
Note that $\mathcal{H}_x$ and $\mathcal{H}_y$ are now related to each other because of the constraints on the interaction of the two mappings.

We then reformalize the learning of latent matching model, referred to as LMM for short, as the following optimization problem:
\begin{equation}\label{eq:lmm}
\begin{split}
  \arg \min_{L_x, L_y}  -\frac{1}{n}\sum_{i=1}^n x_i^T L_x^T L_y y_i + \sum_{p=1,2} \frac{\theta_p}{2} \| L_x^T L_y \|_p^p \\
  + \sum_{p=1,2} \frac{\lambda_p}{2} \| L_x\|_p^p + \sum_{p=1,2} \frac{\rho_p}{2} \|L_y\|_p^p,
\end{split}
\end{equation}
where $\theta_p$, $\lambda_p$ and $\rho_p$ are the hyper-parameters for regularization.

\subsection{Learning Algorithms}\label{sec:learning-algorithm}

In general, there is no guarantee that a global optimal solution of (\ref{eq:lmm}) exists, and thus we employ a greedy algorithm to conduct the optimization. Let $F$ denote the corresponding objective function. The matching term in $F$ can be reformulated as:
\[
\frac{1}{n}\sum_{i=1}^n x_i^T L_x^T L_y y_i = \sum_{u,v} c_{u,v}l_{xu}^T l_{yv},
\]
where $c_{u,v}$ is the $(u,v)$-th element of the empirical cross-covariance matrix $C=\frac{1}{n}\sum_{i=1}^n x_i y_i^T$.

For simplicity, in the following derivation, let us only consider the use of $\ell_2$ regularization, i.e., set $\theta_1=\lambda_1=\rho_1=0$.

For a fixed $L_y$, the derivative of $F$ with respect to $l_{xu}$ is
\[
\frac{\partial F}{\partial l_{xu}} = -\sum_v c_{u,v} l_{yv}^T + \theta_2 \sum_v (l_{xu}^T l_{yv}) l_{yv}^T  + \lambda_2 l_{xu}^T,
\]
and for a fixed $L_x$, the derivative of $F$ with respect to $l_{yv}$ is
\[
\frac{\partial F}{\partial l_{yv}} = - \sum_u c_{u, v} l_{xu}^T + \theta_2 \sum_u (l_{xu}^T l_{yv}) l_{xu}^T + \rho_2 l_{yv}^T.
\]

By setting the derivatives to zeros, the optimal values of $l_{xu}$ and $l_{yv}$ can be solved as:
\begin{equation}\label{eq:lmm-cd}
\begin{split}
l_{xu}^* = \left( \theta_2 \sum_v l_{yv} l_{yv}^T + \lambda_2 I \right)^{-1} \left(\sum_v c_{u, v} l_{yv}\right),\\
l_{yv}^* = \left( \theta_2 \sum_u l_{xu} l_{xu}^T + \rho_2 I \right)^{-1} \left(\sum_u c_{u, v} l_{xu}\right).
\end{split}
\end{equation}
The parameters of $L_x$ and $L_y$ are updated alternatively until convergence.
Algorithm \ref{alg:lmm-cd} shows the main procedure of the coordinate descent algorithm for LMM.

\begin{algorithm}[h]
\caption{Coordinate Descent Algorithm for Latent Matching Model}\label{alg:lmm-cd}
\SetAlgoLined
1. Input: $C$, $\theta_2$, $\lambda_2$, $\rho_2$, $T$.

2. Initialization: $t \leftarrow 0$, random matrices $L_x^{(0)}$ and $L_y^{(0)}$.

\While {not converge and $t \leqslant T $}{
    Compute $A_{x} = \theta_2 L_x^{(t)} (L_x^{(t)})^T + \lambda_2 I$ and its inverse $A_{x}^{-1}$.\\
    Compute $A_{y} = \theta_2 L_y^{(t)} (L_y^{(t)})^T + \rho_2 I$ and its inverse $A_{y}^{-1}$. \\
    Compute $B_{x} = L_x^{(t)} C$. \\
    Compute $B_{y} = C (L_y^{(t)})^T$. \\
    \For{$u=1:d_x$ }{
        Select $u$-th row of $B_{y}$ as $b_{yu}^T$,\\
        Compute $l_{xu}^{(t+1)} = A_{y}^{-1} b_{yu}$.\\
    }
    \For {$v=1:d_y$}{
        Select $v$-th column of $B_{x}$ as $b_{xv}$,\\
        Compute $l_{yv}^{(t+1)} = A_{x}^{-1} b_{xv}$.\\
    }
}
\end{algorithm}

It should be noted that since the parameters are directly calculated, the convergence rate is fast for the coordinate descent algorithm.
However, the calculations at each step in Algorithm \ref{alg:lmm-cd} involve inversion of two $d$-dimension matrices, which could become a computation bottleneck when the dimension of latent space is high. Therefore, we also provide a gradient descent algorithm for LMM as an alternative, specifically for the case of high-dimensional latent space. In this algorithm, $l_{xu}$ and $l_{yv}$ are updated as
\begin{equation}\label{eq:lmm-gd}
\begin{split}
l'_{xu} = l_{xu} + \gamma\left( \sum_v c_{u,v} l_{yv} - \theta_2 \sum_v l_{yv} l_{yv}^T l_{xu} - \lambda_2 l_{xu} \right),\\
l'_{yv} = l_{yv} + \gamma\left( \sum_u c_{u, v} l_{xu} - \theta_2 \sum_u l_{xu} l_{xu}^T l_{yv}  - \rho_2 l_{yv} \right),
\end{split}
\end{equation}
where $\gamma$ is the learning rate. The gradient descent algorithm has less computation at each step but generally needs more iterations to converge. Therefore, one always needs to consider selecting a more suitable optimization method in a specific situation.

When Algorithm \ref{alg:lmm-cd} is applied to RMLS (by letting $\theta_2=0$), the updates of parameters in each iteration become $L_x^{(t+2)}=L_x^{(t)}(\lambda_2\rho_2)^{-1}CC^T$ and $L_y^{(t+2)}=L_y^{(t)}(\lambda_2\rho_2)^{-1}C^TC.$ They are equivalent to conducting power iteration on each row of $L_x$ and $L_y$ independently. Consequently, all rows of $L_x$ will converge to the eigenvector (with the largest eigenvalue) of the matrix $(\lambda_2\rho_2)^{-1}CC^T$, and so will be all rows of $L_y$. Thus, the optimal parameters $L_x^*$ and $L_y^*$ are both matrices of rank one. This justifies the necessity of regularization on the matching matrix $L_x^TL_y$.

With the learned mapping matrices $L_x$ and $L_y$, we can calculate the matching score as in (\ref{eq:matching}). In search, we combine the latent matching score and term matching score, in a similar way as that of Bai et al.'s work~\cite{bai2009supervised}:
\begin{equation}\label{eq:matching-search}
\mathrm{score^{IR}}(x,y) = x^T (L_x^T L_y + I) y =  x^T L_x^T L_y y + x^T y.
\end{equation}
$x^T y$ is equivalent to the matching score in traditional VSM model. Note that formula (\ref{eq:matching-search}) requires that the queries and documents share the same original space, i.e. $\mathcal{X}=\mathcal{Y}$, which can be easily achieved by merging their vocabularies.

\section{Incorporating Semantic Knowledge}\label{sec:lmm-knowledge}

\subsection{Problem Formulation}

A latent matching model trained with the method described in the previous section can perform well for head queries and documents, since it can capture the matching information from click-through data. However, for tail queries and documents, there is not enough click-through data, and it is almost impossible to accurately learn the matching relations between them. To alleviate this problem, we propose incorporating semantic knowledge of synonyms and semantic categories into the learning of the latent matching model.

Without loss of generality, we assume that in one space  the semantic knowledge is represented as a set of pairs of similar objects (e.g., words or tags), denoted as $\{w_i^{(1)}, w_i^{(2)}, s_i\}_{i=1}^{m}$, where $w_i^{(1)}$ and $w_i^{(2)}$ represent the term vectors of the objects, and $s_i$ is a scalar representing their weight.
Therefore, the matching degrees of the pairs become
\begin{equation}\label{eq:knowledge-formalization}
\sum_{i=1}^{m} s_i\; (w_{i}^{(1)})^T L^T L w_{i}^{(2)}.
\end{equation}

We extend the latent matching model (\ref{eq:lmm}) by incorporating the above `regularization' term, for the two spaces $\mathcal{X}$ and $\mathcal{Y}$ respectively, into the objective function of learning:
\begin{equation}\label{eq:lmm-knowledge}
\begin{split}
  \arg \min_{L_x, L_y}  -\frac{1}{n}\sum_{i=1}^n x_i^T L_x^T L_y y_i -\frac{\alpha}{m_x} \sum_{i=1}^{m_x} s_{x,i}\; (w_{x,i}^{(1)})^T L_x^T L_x w_{x,i}^{(2)}\\
  - \frac{\beta}{m_y} \sum_{i=1}^{m_y} s_{y,i}\; (w_{y,i}^{(1)})^T L_y^T L_y w_{y,i}^{(2)} + \sum_{p=1,2} \frac{\theta_p}{2} \| L_x^T L_y \|_p^p \\
  + \sum_{p=1,2} \frac{\lambda_p}{2} \| L_x\|_p^p + \sum_{p=1,2} \frac{\rho_p}{2} \|L_y\|_p^p.
\end{split}
\end{equation}
The hyper-parameters $\alpha$ and $\beta$ control the importance of semantic knowledge from the two spaces.

Similarly, coordinate descent can be employed to solve the optimization (\ref{eq:lmm-knowledge}).
The optimal values of $l_{xu}$ and $l_{yv}$ are then given by
\begin{equation}\label{eq:lmm-knowledge-cd}
\begin{split}
l_{xu}^* = \left( \theta_2 \sum_v l_{yv} l_{yv}^T + \lambda_2 I \right)^{-1} \left(\sum_v c_{u, v} l_{yv} + \alpha \sum_v r_{x,u,v} l_{xv}\right),\\
l_{yv}^* = \left( \theta_2 \sum_u l_{xu} L_{xu}^T + \rho_2 I \right)^{-1} \left(\sum_u c_{u, v} l_{xu} + \beta \sum_u r_{y,u,v} l_{yu}\right),
\end{split}
\end{equation}
where $r_{x,u,v}$ and $r_{y,u,v}$ denote the $(u,v)$-th elements of the empirical covariance matrices $R_x$ and $R_y$ respectively, where
\[
R_x=\frac{1}{m_x}\sum_{i=1}^{m_x} s_{x,i} \; w_{x,i}^{(1)} (w_{x,i}^{(2)})^T \textrm{, }
R_y=\frac{1}{m_y}\sum_{i=1}^{m_y} s_{y,i} \; w_{y,i}^{(1)} (w_{y,i}^{(2)})^T.
\]
Algorithm \ref{alg:lmm-knowledge-cd} shows the procedure of the coordinate descent algorithm for latent matching model with semantic knowledge. When compared with Algorithm~\ref{alg:lmm-cd}, Algorithm \ref{alg:lmm-knowledge-cd} clearly shows how semantic knowledge takes effect in training: the right-hand side matrices of the linear equations ($B_x$ and $B_y$) are corrected by the covariance matrices derived from semantic knowledge, while the left-hand side matrices ($A_x$ and $A_y$) stay invariant.

\begin{algorithm}[h]
\SetAlgoLined
\caption{Coordinate Descent Algorithm for Latent Matching Model with Semantic Knowledge}\label{alg:lmm-knowledge-cd}
1. Input: $C$, $R_x$, $R_y$, $\alpha$, $\beta$, $\theta$, $\lambda_2$, $\rho_2$, $T$.

2. Initialization: $t \leftarrow 0$, random matrices $L_x^{(0)}$ and $L_y^{(0)}$.

\While {not converge and $t \leqslant T $}{
    Compute $A_{x} = \theta_2 L_x^{(t)} (L_x^{(t)})^T + \lambda_2 I$ and its inverse $A_{x}^{-1}$ .\\
    Compute $A_{y} = \theta_2 L_y^{(t)} (L_y^{(t)})^T + \rho_2 I$ and its inverse $A_{y}^{-1}$. \\
    Compute $B_{x} = L_x^{(t)} C + \beta L_y^{(t)}R_y $. \\
    Compute $B_{y} = C (L_y^{(t)})^T + \alpha R_x (L_x^{(t)})^T$. \\
    \For{$u=1:d_x$ }{
        Select $u$-th row of $B_{y}$ as $b_{yu}^T$,\\
        Compute $l_{xu}^{(t+1)} = A_{y}^{-1} b_{yu}$.\\
    }
    \For {$v=1:d_y$}{
        Select $v$-th row of $B_{x}$ as $b_{xv}^T$,\\
        Compute $l_{yv}^{(t+1)} = A_{x}^{-1} b_{xv}$.\\
    }
}
\end{algorithm}

An alternative algorithm using gradient descent can be obtained by updating $l_{xu}$ and $l_{yv}$ as
\begin{equation}\label{eq:lmm-knowledge-gd}
\begin{split}
l'_{xu} = l_{xu} + \gamma\left( \sum_v c_{u,v} l_{yv} + \alpha \sum_v r_{x,u,v} l_{xv} \right) \\
    - \gamma \left( \theta_2 \sum_v l_{yv} l_{yv}^T l_{xu} + \lambda_2 l_{xu} \right),\\
l'_{yv} = l_{yv} + \gamma \left( \sum_u c_{u, v} l_{xu} + \beta \sum_u r_{y,u,v} l_{yu} \right) \\
    - \gamma \left( \theta_2 \sum_u l_{xu} l_{xu}^T l_{yv}  + \rho_2 l_{yv} \right),
\end{split}
\end{equation}
where $\gamma$ is the learning rate.

\subsection{Acquisition of Semantic Knowledge}\label{sec:semantic-knowledge-aquisition}

The question next is how to acquire and represent the semantic knowledge in our method. We can consider multiple ways of creating a semantic knowledge base, either manually or automatically. In this section, we briefly describe two methods to acquire and represent semantic knowledge.

Synonyms are obviously useful semantic knowledge for our matching task. A general dictionary of synonyms such as WordNet is usually not suitable for a real-world setting, however. The reason is that synonyms usually heavily depend on domains. Here we adopt an algorithm for mining synonym pairs by exploiting click-through data. Specifically, we first try to find clusters of queries from a click-through bipartite graph (cf., \cite{jiang2013mining}). Queries in one cluster are regarded as synonyms. Next, for each cluster, we extract pairs of terms sharing the same context as candidates of synonym pairs (cf., \cite{lamping2009determining}). Here the context refers to the surrounding text of a term in the query. For example, given a query ``download 2048 apk'', the context for term ``2048'' is ``download * apk'', where `*' is the wildcard character. Then we go through all the clusters and count the numbers of occurrences (called support) for all the candidate pairs. The candidate pairs with support above a certain threshold are chosen as synonym pairs. Algorithm \ref{alg:synonym-mining} shows the detailed procedure.

\begin{algorithm}[h]
\caption{Synonyms Mining Algorithm on Click Bipartite Graph}\label{alg:synonym-mining}
\SetAlgoLined
0. Notation: $Q$: query set, $D$: document set, $C$: click set, $q$: query, $d$: document, $t$: term.

1. Input: click bipartite graph $G=(Q, D, C)$.

2. Initialization: dictionary of candidate synonym pairs $S=[\ ]$.

\For {$d$ in $D$ }{
    Collect $ Q_d = \{q | (q, d) \in C\}$.\\
    Init $T=\{\ \}$.\\
    \For {$q$ in $Q_d$} {
        \For {$t$ in $q$}{
            Extract context $c_t$ of $t$ in $q$ \\
            Add ($t$, $c_t$) to $T$ \\
        }
    }
    Find $P_d = \{(t_i, t_j) | c_{t_i} = c_{t_j}, (t_i, c_{t_i}) \in T, (t_j, c_{t_j}) \in T\}$\\
    \For {$(t_i, t_j)$ in $P_d$}{
        \eIf {$(t_i, t_j)$ not in $S$ }{
            Add $(t_i, t_j)$ to $S$ and set $S[(t_i, t_j)] = 1$
            }{
            Set $S[(t_i, t_j)] = S[(t_i, t_j)] + 1$
            }
    }
}
3. Sort $S$ by value in descending order.\\
4. Return top $K$ pairs of $S$ as the synonym pairs.
\end{algorithm}

We denote the set of mined synonym pairs as $\{(w_{x,i}^{(1)}, w_{x,i}^{(2)}, s_{x,i})\}$, where $w_{x,i}^{(1)}$ and $w_{x,i}^{(2)}$ are the $i$-th pair of synonyms. $s_{x,i}$ is the corresponding weight for the pair, which is computed as the logistic transformation of the support. Following (\ref{eq:knowledge-formalization}), the knowledge about the synonym set for the query domain ($\mathcal{X}$) is formalized as $\sum_{i}s_{x,i}\; (w_{x,i}^{(1)})^T L_x^T L_x w_{x,i}^{(2)}$ in the optimization function.
It should be noted that although using the same click-through data, the synonym mining algorithm makes use of a different part of the data, i.e., click-partite graph, and thus can provide complementary information to the learning algorithm of latent matching model.

In addition to synonyms, we also utilize categories or tags in a taxonomy as semantic knowledge for the document domain. For example, in our experiment of the mobile app search, apps are given various tags by users. An app named ``tiny racing car'' is tagged ``action, mario, racing, sports, auto, adventure, racing track''. For each tag, we have a list of associated documents. We represent the title of each document as a tf-idf vector and calculate the average vector of the tf-idf vectors for each tag.
We select the top $k$ terms in the average vector and view them as the relevant terms to the tag.
A set of `tag-term' pairs is then obtained from all the tags and their relevant terms, and it is denoted as $\{(w_{y,i}, w_{y,ij}, s_{y,ij})\}$ ,
where $w_{y,i}$ is the $i$-th tag, and $w_{y,ij}$ is the $j$-th relevant term to the $i$-th tag, and $s_{y,ij}$ is the corresponding average tf-idf value.
Algorithm \ref{alg:tags-mining} shows the detailed procedure. We can formalize the knowledge for the document domain ($\mathcal{Y}$) as $\sum_{i}\sum_{j}s_{y,ij}\; (w_{y,i}^{(1)})^T L_y^T L_y w_{y,ij}^{(2)}$ in the objective function of learning of latent matching model.\\

\begin{algorithm}[h]
\caption{Tag-Term Mining Algorithm on App Data}\label{alg:tags-mining}
\SetAlgoLined
0. Notation: $T$: tag set, $D$: document set, $t$: tag, $d$: document, $w$: term, $D[t]$: set of documents with tag $t$.

1. Input: set of documents and set of tags $(D, T)$.

2. Initialization: set of candidate term-tag pairs $S=[\ ]$.

    \For {$t$ in $T$}{
        Init term vector $v$ as zeros\\
        \For {$d$ in $D[t]$} {
            Compute tf-idf vector of document $d$ as $v_d$\\
            $v$ = $v$ + $v_d$\\
        }
        $\bar{v} = \frac{v}{\|D[t]\|}$\\
        Sort $\bar{v}$ and find top $k$ term indices as $W[t]$\\
        \For {$i$ in $W[t]$} {
            Set $S[(t, w_i)] = \bar{v}[i]$
        }
    }
    3. Return S as set of tag-term pairs.\\
\end{algorithm}

\subsection{Relation with Latent Factor Models}
Latent factor models are state-of-the-art methods for collaborative filtering. They are basically models for matching users and items in latent spaces. In fact, our latent matching model shares similarities with the latent factor models. For example, Regularized Single Value Decomposition (RSVD) ~\cite{paterek2007improving,srebro2004maximum}, a widely used matrix factorization method for collaborative filtering, can be formalized as the following optimization problem:
\begin{equation}\label{eq:rsvd}
\arg \min_{U, V} \sum_{i,j} (r_{i,j} - u_i^T v_j)^2 + \lambda_1 \| U \|_2^2 + \lambda_2 \| V \|_2^2,
\end{equation}
where $u_i$ and $v_j$ are the column vectors of the user matrix $U \in \mathbb{R}^{d \times d_u}$ and the item matrix $V \in \mathbb{R}^{d \times d_v}$.
By expanding the squared loss
\[
\sum_{i,j} (r_{i,j} - u_i^T v_j)^2 = \sum_{i,j} r_{i,j}^2 - 2\sum_{i,j} r_{i,j}u_i^T v_j + \sum_{i,j} (u_i^T v_j)^2,
\]
we can see that the first term is a constant, the second term is analogous to the matching degree in LMM, and the third term corresponds to the $\ell_2$ regularization on the matching matrix. Actually, LMM subsumes RSVD as a special case and is more general and flexible.

Ma~\cite{Ma:2013ae} propose incorporating user and item social information in RSVD, by adding  regularization terms into the objective function. Their work is similar to ours in the sense that additional knowledge is exploited to enhance the accuracy of the learned model. However, there are clear distinctions. First, the formulations are different. We adopt inner product as similarity measure, which seem to be more appropriate for search. Second, the applications are different and our work seems to be the first attempt for search.

\subsection{Parallelization of Learning}
It is often necessary to train a latent matching model with a huge amount of training data (e.g., click-through logs) and from two spaces of very high dimensionalities (e.g., a very large vocabulary of queries and documents). We also consider parallelization of the learning algorithms to deal with the challenge.

The coordinate descent algorithm described in Algorithm \ref{alg:lmm-cd} and Algorithm \ref{alg:lmm-knowledge-cd} can be parallelized. First, for initialization, one can calculate the empirical cross-variance matrix $C$ by randomly partitioning the training data into subsets and conducting calculation on each of the subsets, and summing up the results. Second, at each iteration of coordinate descent, all the calculations are basically matrix/vector multiplications, summations, and inversions. One can exploit existing parallelization schemes and tools for the computations. Specifically in our experiment, the mapping matrices $L_x$ and $L_y$ are partitioned by column and the sub-matrices are calculated simultaneously on multiple processors. For matrix inversion, we employ the parallelized LU-decomposition and then forward and backward substitution to solve the linear system. Furthermore, the computation of inversions of $A_x$ and $A_y$ in Algorithm \ref{alg:lmm-cd} and \ref{alg:lmm-knowledge-cd} can be omitted. Instead we directly solve the linear system with multiple right-hand sides $A_y L_x = B_y$ and $A_x L_y = B_x$, by applying the parallelized LU-decomposition.

The gradient descent can also be parallelized in a similar way. We omit the details.

\section{Experiments}\label{sec:experiments}

We have conducted experiments to test the matching performances of latent matching models on relevance ranking in search. The training and test data are click-through logs obtained from a mobile app search engine, where each app is deemed as a document.
We first compare the performance of the LMM model with BM25, PLS, RMLS, on the same datasets. After that, we test the performance of the LMM model augmented with semantic knowledge. The semantic knowledge includes synonym pairs mined from the click-through logs, and semantic tags assigned to the apps.

\subsection{Experimental Setup}
We take app search as example and use data from an app search engine. Each app is represented by its title and description and can be viewed as a document. Click-through logs at the search engine are collected and processed.
We create two datasets from the click-through logs, one containing one week data and the other containing one month data. Table \ref{tab:data-statistics} reports some statistics of the two datasets. It should be noted that standard datasets in IR do not have such a large amount of labeling information, so as to train the latent matching model.

Each dataset consists of query-document pairs and their associated clicks, where a query and a document are represented by a term-frequency vector and a tf-idf vector of the title, respectively.
 The queries and documents are regarded as heterogeneous data in two different spaces, because queries and documents have different characteristics.

\begin{table}
\centering
\caption{Statistics of two training datasets}\label{tab:data-statistics}
\begin{tabular}{c c c c}
\hline
	& \#clicks & \#queries & \#apps\\
\hline
\hline
one-week click-through data &1,020,854 &190,486&110,757\\
\hline
one-month click-through data &3,441,768&534,939&192,026\\
\hline
\end{tabular}
\end{table}

In addition, we randomly sample two sets of 200 queries from a time period different from that of training datasets, and take them as two test datasets.

Each test set is composed of 100 head queries and 100 tail queries, according to the frequencies of them.
In the following sub-sections, performance on the whole random test set as well as the head and tail subsets will be reported.
For each query in the test sets, we collect top 20 apps retrieved by the app search engine and then label the query-app pairs at four levels of matching: Excellent, Good, Fair, and Bad.

As evaluation measures, Normalized Discounted Cumulative Gain (NDCG) at positions 1, 3, 5, 10 are used.
We choose the conventional IR model of BM25 (with the parameters tuned for best performance in the training set), and two latent matching models of PLS (Partial Least Square) and RMLS (Regularized Mapping to Latent Structures) as the baseline methods. Our basic model is denoted as LMM (Latent Matching Model) and our augmented models are denoted as LMM-X where X stands for the type of incorporated semantic knowledge.

All the experiments are conducted on a Linux server with 24-core Intel Xeon 1.9GHz CPUs and 192GB RAM. In model training, we employ the OPENBLAS tool\footnote{http://www.openblas.net} for matrix computation. In order to leverage the multi-core machine, we extensively use parallel-process techniques.

\subsection{Experimental results}
\subsubsection{Latent Matching Model}
We conduct a series of experiments to test the performances of LMM, LMM-X and the baseline models. For RMLS, LMM, and LMM-X, the results with latent dimensionalities of 100 and 500 are reported. For PLS, only the performance with latent dimensionality of 100 is reported, due to its scalability limitation.

Table \ref{tab:one-week-lmm-random-queries} ,\ref{tab:one-week-lmm-head-queries} and \ref{tab:one-week-lmm-tail-queries} report the performances of the models trained using one-week click-through data, evaluated on the test tests: random queries, head queries and tail queries respectively. From the results, we can see that:
(1) all the latent matching models significantly outperform the conventional BM25 model in terms of all evaluation measures;
(2) among the latent space models with the same dimension, LMM achieves the best performances in many cases. The improvements of LMM over BM25 and RMLS are statistically significant (paired t-test, p-value<0.05);
(3) the improvements of LMM over the other baseline models are larger on tail queries than on head queries, which indicates that LMM can really enhance matching performance for tail queries;
(4) for LMM, the performance increases as the dimensionality of latent space increases.
Note that PLS requires SVD and thus becomes practically intractable when the dimension is large.
In that sense, RMLS and LMM exhibit their advantages over PLS on scalability.

\begin{table*}
\centering
  \caption{Matching performance on one week data on Random queries}\label{tab:one-week-lmm-random-queries}
  \begin{tabular}{|c|c c c c|}
    \hline
    Model & \multicolumn{4}{|c|}{NDCG on Random queries}  \\
    (Dimension) & @1 & @3 & @5 & @10 \\
    \hline
    BM25 & 0.687 & 0.700 & 0.707 & 0.741\\
    PLS(100) & 0.715 & 0.733 & 0.738 & 0.767  \\
    RMLS(100) & 0.697 & 0.727 &0.732  & 0.765   \\
    LMM(100) & 0.713 & 0.727 & 0.741 & 0.771  \\
    RMLS(500) & 0.709 & 0.720 & 0.731 & 0.762 \\
    LMM(500) &\textbf{0.727} &\textbf{ 0.737} & \textbf{0.738}& \textbf{0.772} \\
    \hline
  \end{tabular}
\end{table*}

\begin{table*}
\centering
  \caption{Matching performance on one week data on Head queries}\label{tab:one-week-lmm-head-queries}
  \begin{tabular}{|c|c c c c|}
    \hline
    Model & \multicolumn{4}{|c|}{NDCG on Head queries} \\
    (Dimension) & @1 & @3 & @5 & @10 \\
    \hline
    BM25 & 0.729 & 0.754 & 0.758 & 0.786  \\
    PLS(100)  & 0.756 & 0.780 & 0.787 & 0.809 \\
    RMLS(100)  & 0.740 & 0.767 & 0.772 & 0.801  \\
    LMM(100)  & 0.744 & 0.771 & 0.785 & 0.813  \\
    RMLS(500)  & 0.742 & 0.765 & 0.777 & 0.805  \\
    LMM(500)  & \textbf{0.766} &\textbf{0.783} & \textbf{0.787} &\textbf{0.812} \\
    \hline
  \end{tabular}
\end{table*}

\begin{table*}
\centering
  \caption{Matching performance on one week data on Tail queries}\label{tab:one-week-lmm-tail-queries}
  \begin{tabular}{|c|c c c c|}
    \hline
    Model &  \multicolumn{4}{|c|}{NDCG on Tail queries} \\
    (Dimension) & @1 & @3 & @5 & @10\\
    \hline
    BM25 &0.645 & 0.645 &0.656  & 0.696 \\
    PLS(100) &  0.675 & 0.686 & 0.689 & 0.726 \\
    RMLS(100) &0.653& 0.686 & 0.692 & 0.729  \\
    LMM(100) &  0.681 & 0.684 & \textbf{0.697} & 0.729 \\
    RMLS(500) &  0.677 & 0.674 & 0.686 & 0.719 \\
    LMM(500) &\textbf{0.689} & \textbf{0.690} & 0.688 & \textbf{0.731} \\
    \hline
  \end{tabular}
\end{table*}

Table \ref{tab:one-month-lmm-random-queries}, \ref{tab:one-month-lmm-head-queries} and \ref{tab:one-month-lmm-tail-queries} show the comparison results of models trained using one-month click-though data, evaluated on the tested random queries, head queries and tail queries respectively, which follows the same trends as that of one-week data, especially on tail queries.

\begin{table*}
\centering
  \caption{Matching performance on one-month data on Random queries}\label{tab:one-month-lmm-random-queries}
  \begin{tabular}{|c|c c c c|}
    \hline
    Model & \multicolumn{4}{|c|}{NDCG on Random queries} \\
    (Dimension) & @1 & @3 & @5 & @10\\
    \hline
    BM25 & 0.644& 0.681 & 0.714 & 0.740  \\
    PLS(100) & 0.692 & 0.727 & 0.749 & 0.772  \\
    RMLS(100) & 0.668 & 0.703 & 0.727 & 0.752 \\
    LMM(100) & 0.692 & \textbf{0.733} & \textbf{0.751} & 0.775  \\
    RMLS(500) & 0.687 & 0.725 & 0.745& 0.774 \\
    LMM(500) & \textbf{0.704}& 0.730 & 0.749 & \textbf{0.780} \\
    \hline
  \end{tabular}
\end{table*}

\begin{table*}
\centering
  \caption{Matching performance on one-month data on Head queries}\label{tab:one-month-lmm-head-queries}
  \begin{tabular}{|c|c c c c|}
    \hline
    Model &  \multicolumn{4}{|c|}{NDCG on Head queries}  \\
    (Dimension) & @1 & @3 & @5 & @10 \\
    \hline
    BM25  & 0.721 & 0.738 & 0.756 & 0.771  \\
    PLS(100) & 0.735 & 0.757 & 0.774 & 0.788  \\
    RMLS(100)  & 0.736 & 0.746 & 0.762 & 0.779 \\
    LMM(100) & 0.744 & 0.765 & \textbf{0.779} & 0.793  \\
    RMLS(500)  & \textbf{0.753}& \textbf{0.767} & 0.772 & 0.798 \\
    LMM(500)  & 0.745 & 0.756 & 0.770 & \textbf{0.795}  \\
    \hline
  \end{tabular}
\end{table*}

\begin{table*}
\centering
  \caption{Matching performance on one-month data on Tail queries}\label{tab:one-month-lmm-tail-queries}
  \begin{tabular}{|c|c c c c|}
    \hline
    Model &  \multicolumn{4}{|c|}{NDCG on Tail queries} \\
    (Dimension) & @1 & @3 & @5 & @10 \\
    \hline
    BM25 & 0.567 & 0.624 & 0.672 & 0.710 \\
    PLS(100) &  0.649 & 0.698 & 0.724 & 0.756 \\
    RMLS(100)& 0.600 & 0.660 & 0.693& 0.726 \\
    LMM(100) & 0.640 & 0.700 & 0.724 & 0.758 \\
    RMLS(500) &  0.620 & 0.684 & 0.719 & 0.751 \\
    LMM(500) & \textbf{0.662} & \textbf{0.704} & \textbf{0.729} & \textbf{0.765} \\
    \hline
  \end{tabular}
\end{table*}

\subsubsection{Incorporating Semantic knowledge}
Next, we test the performances of the LMM-X models which incorporate semantic knowledge into the model.
The LMM-X models have the ability to leverage multiple sources of semantic knowledge by adding regularization terms to the objective function. We consider two methods of acquiring and utilizing semantic knowledge.

In the first method we mine and use synonym pairs from the click-through logs, by employing Algorithm \ref{alg:synonym-mining}. In the second method we collect and use over 50,000 tags in the app search engine, by employing Algorithm \ref{alg:tags-mining}.

We conduct experiments using LMM model and the two types of knowledge. We summarize the results in Table \ref{tab:one-week-lmm-knowledge-random-queries}, \ref{tab:one-week-lmm-knowledge-head-queries} and \ref{tab:one-week-lmm-knowledge-tail-queries} for one-week data and Table \ref{tab:one-month-lmm-knowledge-random-queries},\ref{tab:one-month-lmm-knowledge-head-queries} and \ref{tab:one-month-lmm-knowledge-tail-queries} for one-month data evaluated on random queries, head queries and tail queries respectively.  For each training dataset, we first separately train the LMM model augmented with the synonyms dictionary and the tag-term pairs, denoted as LMM-Synonyms and LMM-Tags, respectively. Then we train the LMM model augmented with both types of knowledge, denoted as LMM-Both. From the results we can see:
(1) with knowledge embedded, the performances of the LMM model can be consistently improved;
(2) the improvements of LMM-Both over LMM are statistically significant (paired t-test, p-value<0.05) in terms of most evaluation measures;
(3) more significant improvements are made on tail queries than on head queries;
(4) the improvements of semantic knowledge augmentation are slightly less when the latent dimensionality is high (500) than when it is low (100).

\begin{table*}
\centering
  \caption{Mathcing performance on one-week data on Random queries}\label{tab:one-week-lmm-knowledge-random-queries}
  \begin{tabular}{|c|c c c c|}
    \hline
    Model & \multicolumn{4}{|c|}{NDCG on Random queries} \\
    (Dimension) & @1 & @3 & @5 & @10 \\
    \hline
    LMM(100) & 0.713 & 0.727 & 0.741 & 0.771\\
    LMM-Synonyms(100) & 0.730 & 0.743 & 0.747 & 0.772 \\
    LMM-Tags(100) & 0.727 & 0.746 & 0.747& 0.773  \\
    LMM-Both(100) & 0.735 & 0.750 & 0.752 & 0.772\\
    \hline
    LMM(500) & 0.727 & 0.737 & 0.738 & 0.772 \\
    LMM-Synonyms(500) & 0.743 & 0.749 & 0.758 & 0.781 \\
    LMM-Tags(500) & 0.743 & 0.747 & 0.759 & \textbf{0.783} \\
    LMM-Both(500) & \textbf{0.743} & \textbf{0.750} & \textbf{0.759} & 0.781 \\
    \hline
  \end{tabular}
\end{table*}

\begin{table*}
\centering
  \caption{Mathcing performance on one-week data on Head queries}\label{tab:one-week-lmm-knowledge-head-queries}
  \begin{tabular}{|c|c c c c|}
    \hline
    Model &  \multicolumn{4}{|c|}{NDCG on Head queries}  \\
    (Dimension) & @1 & @3 & @5 & @10\\
    \hline
    LMM(100)  &0.744 & 0.771 & 0.785 & 0.813\\
    LMM-Synonyms(100)  &0.757 & 0.791 & 0.794 & 0.815 \\
    LMM-Tags(100)  & 0.757 &0.789 & 0.796& 0.817 \\
    LMM-Both(100) & 0.762 & 0.798 & 0.799 & 0.815 \\
    \hline
    LMM(500) &0.766 & 0.783 & 0.787 & 0.812  \\
    LMM-Synonyms(500) &0.779 & \textbf{0.795} & \textbf{0.802} & 0.819  \\
    LMM-Tags(500) &0.779 & 0.793 & 0.801 & \textbf{0.820} \\
    LMM-Both(500) &\textbf{0.779} & 0.793 & 0.801 & 0.819 \\
    \hline
  \end{tabular}
\end{table*}

\begin{table*}
\centering
  \caption{Mathcing performance on one-week data Tail queries}\label{tab:one-week-lmm-knowledge-tail-queries}
  \begin{tabular}{|c|c c c c|}
    \hline
    Model & \multicolumn{4}{|c|}{NDCG on Tail queries} \\
    (Dimension) & @1 & @3 & @5 & @10\\
    \hline
    LMM(100) &0.681 &0.684&0.697&0.729\\
    LMM-Synonyms(100) & 0.704 &0.695&0.700 &0.729\\
    LMM-Tags(100) &  0.697&0.699 &0.699&0.728 \\
    LMM-Both(100) & \textbf{0.709} &0.702&0.705 &0.729\\
    \hline
    LMM(500)  & 0.689&0.690&0.688 &0.731 \\
    LMM-Synonyms(500) &0.707 &0.703&0.714 &0.743 \\
    LMM-Tags(500) &0.707 &0.702&0.716 &\textbf{0.745}\\
    LMM-Both(500) &0.707 &\textbf{0.708}&\textbf{0.718} &0.743\\
    \hline
  \end{tabular}
\end{table*}

\begin{table*}
\centering
  \caption{Matching performance on one-month data on Random queries}\label{tab:one-month-lmm-knowledge-random-queries}
 \begin{tabular}{|c|c c c c|}
    \hline
    Model & \multicolumn{4}{|c|}{NDCG on Random queries} \\
    (Dimension) & @1 & @3 & @5 & @10 \\
    \hline
    LMM(100) & 0.692 & 0.727& 0.749 & 0.772 \\
    LMM-Synonyms(100) & 0.708 &0.738 & 0.749 & 0.780 \\
    LMM-Tags(100) & 0.707 & 0.734 & 0.750 & 0.779 \\
    LMM-Both(100) & 0.715 & 0.739 & 0.745 & 0.779 \\
    \hline
    LMM(500) & 0.704 & 0.730 & 0.749 & 0.780 \\
    LMM-Synonyms(500) & 0.719 & 0.741 & \textbf{0.762} & \textbf{0.783} \\
    LMM-Tags(500) & 0.719 & 0.741 & 0.762 & 0.781\\
    LMM-Both(500) & \textbf{0.721} & \textbf{0.745} & 0.761 & 0.782 \\
    \hline
  \end{tabular}
\end{table*}

\begin{table*}
\centering
  \caption{Matching performance on one-month data on Head queries}\label{tab:one-month-lmm-knowledge-head-queries}
 \begin{tabular}{|c|c c c c|c c c c| c c c c |}
    \hline
    Model & \multicolumn{4}{|c|}{NDCG on Head queries}  \\
    (Dimension) & @1 & @3 & @5 & @10 \\
    \hline
    LMM(100)  &0.735 & 0.757 & 0.774 & 0.788 \\
    LMM-Synonyms(100) &0.741 & 0.771 & 0.770 & 0.795 \\
    LMM-Tags(100)  &0.738 & 0.760 &0.767& 0.795\\
    LMM-Both(100) &0.738 & 0.760 & 0.767 & 0.795 \\
    \hline
    LMM(500) &0.745& 0.756 & 0.770 & \textbf{0.795} \\
    LMM-Synonyms(500) &\textbf{0.752} & 0.761 & 0.775 & 0.793 \\
    LMM-Tags(500) &0.752 & 0.759 & \textbf{0.778} & 0.794  \\
    LMM-Both(500)  &0.751 & \textbf{0.763} & 0.777 & 0.793 \\
    \hline
  \end{tabular}
\end{table*}

\begin{table*}
\centering
  \caption{Matching performance on one-month data on Tail queries}\label{tab:one-month-lmm-knowledge-tail-queries}
 \begin{tabular}{|c|c c c c|}
    \hline
    Model & \multicolumn{4}{|c|}{NDCG on Tail queries} \\
    (Dimension) & @1 & @3 & @5 & @10\\
    \hline
    LMM(100) & 0.649 &0.698&0.724 &0.756\\
    LMM-Synonyms(100) &0.676 &0.705&0.729 &0.765\\
    LMM-Tags(100)  & 0.676 &0.708 &0.733&0.763 \\
    LMM-Both(100) &0.676 &0.708&0.733 &0.760\\
    \hline
    LMM(500) &0.662 &0.704&0.729 &0.765 \\
    LMM-Synonyms(500)  &0.686 &0.723&\textbf{0.748}&\textbf{0.773}\\
    LMM-Tags(500) &0.686 &0.723&0.746 &0.769 \\
    LMM-Both(500) & \textbf{0.691} &\textbf{0.728}&0.745 &0.771\\
    \hline
  \end{tabular}
\end{table*}

\begin{figure*}[!t]
\centering
          \subfigure[LMM]
          {
          \begin{minipage}[b]{\textwidth}
            \centering
            \label{fig:mml-mds}
            \includegraphics[angle=-90,width=0.7\textwidth]{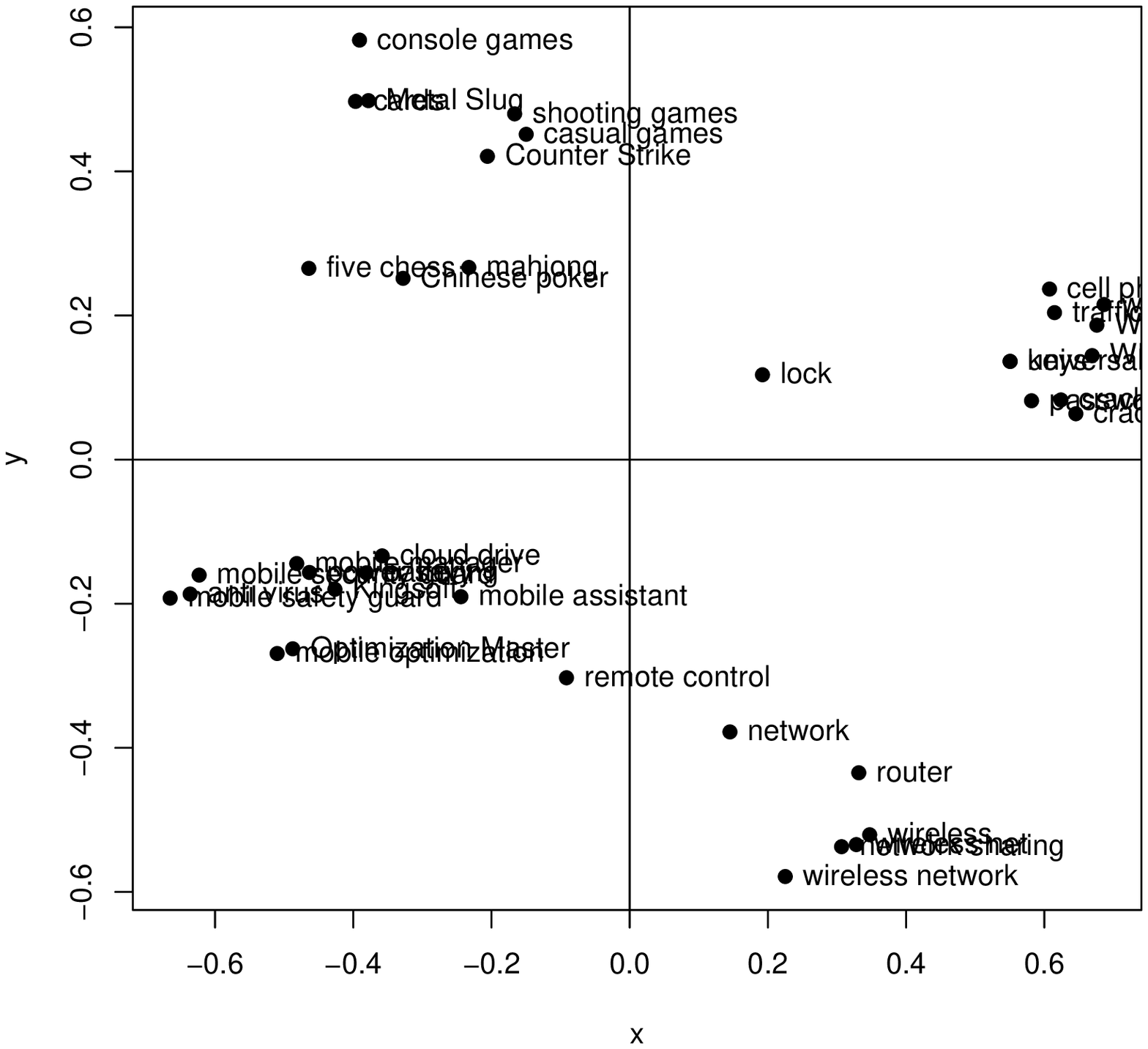}
          \end{minipage}
          }
          \hfill
          \subfigure[LMM-Synonyms]
          {
	\begin{minipage}[b]{\textwidth}
            \centering
	       \label{fig:mml-mds-synonyms}
           \includegraphics[angle=-90,width=0.7\textwidth]{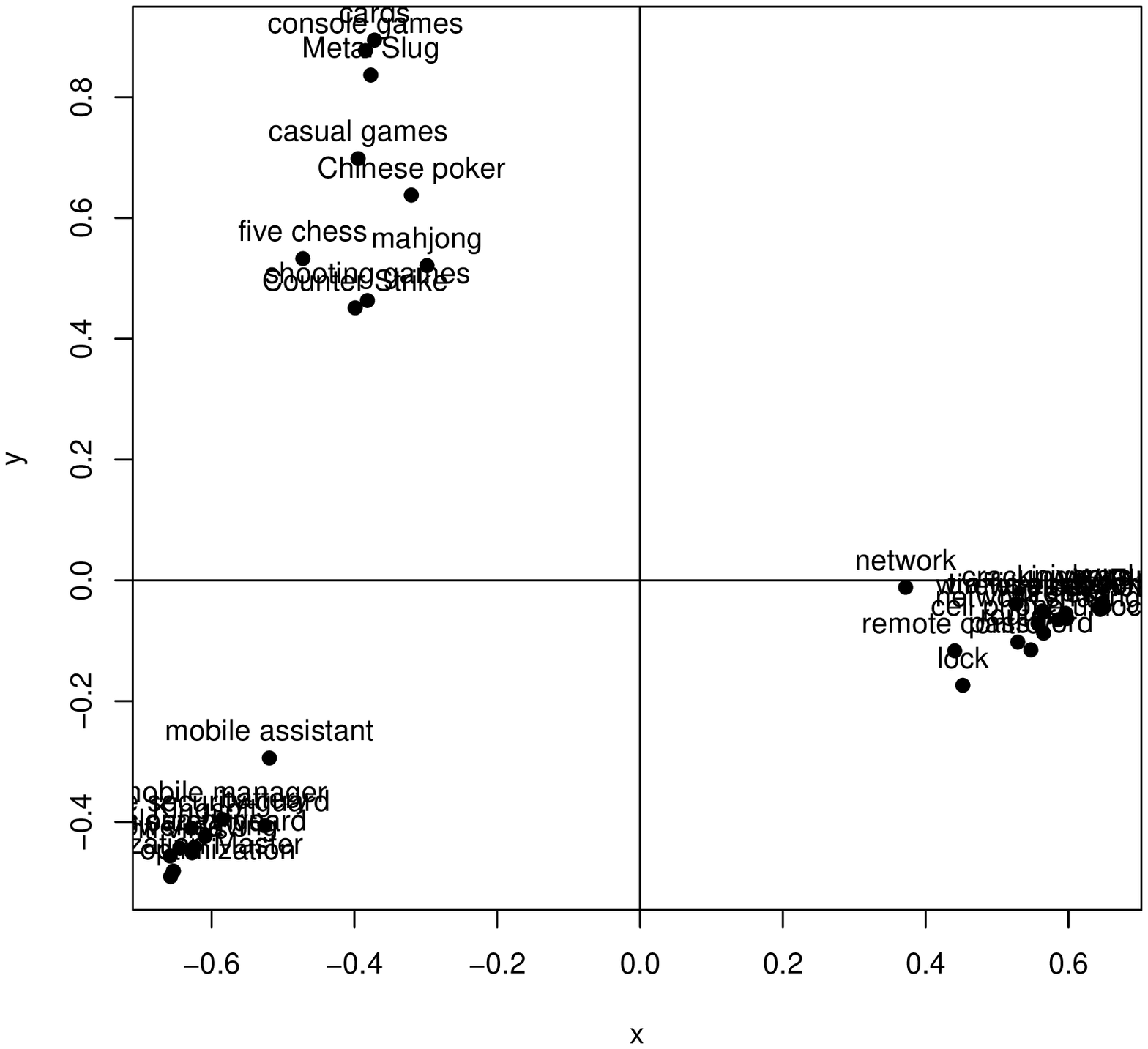}
          \end{minipage}
          }
  \caption{Representations of query words in latent space.}\label{fig:mml-mds-compare}
\end{figure*}

We investigate the latent spaces of LMMs learned with and without incorporating synonym dictionary. The latent representations of some randomly selected words are plotted on a 2-D graph using the multidimensional scaling technique, in Figure~\ref{fig:mml-mds-compare}. By comparing the distributions of words in Figure~\ref{fig:mml-mds} and Figure~\ref{fig:mml-mds-synonyms}, we can clearly see that similar words are clustered closer in LMM-Synonyms than in LMM. This clearly indicates that knowledge about synonyms can be effectively incorporated into LMM-Synonyms and thus the model can further improve matching. For the latent spaces of LMMs learned with and without incorporating category tags, we observe a similar phenomenon.\\

\subsubsection{Parameter Setting}
\begin{figure}
  \centering
  \includegraphics[scale=0.5]{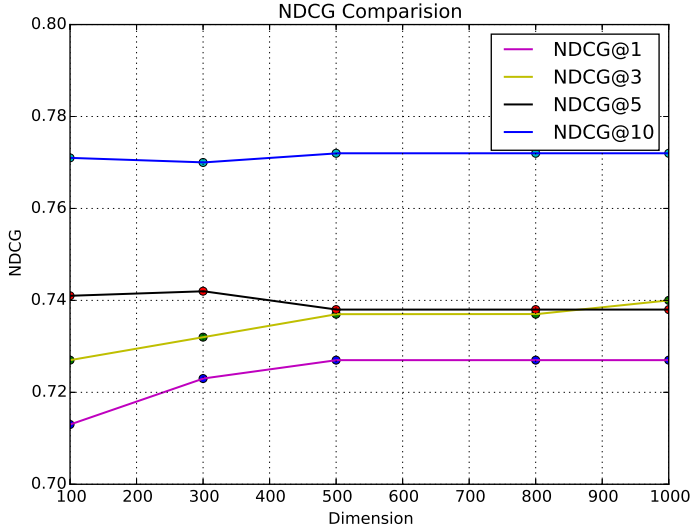}\\
  \caption{NDCG scores versus dimensionalities}\label{fig:parameter-dimension}
\end{figure}

We investigate the effect of hyper-parameters on the learning. The dimension of the latent space is an important parameter. We set the latent dimension in the range of \{100, 300, 500, 800, 1000\}, and show the matching performances on one-week data in Figure \ref{fig:parameter-dimension}. From the results, we can see that when dimensionality increases, the performance generally improves, but the gain gradually becomes smaller, especially after dimensionality exceeds 500. On the other hand, when dimensionally increases, the cost of training will also become high. In the current case, the dimensionality of 500 represents a good trade-off between effectiveness and efficiency.

The regularization hyper-parameters ($\theta$, $\alpha$, $\beta$, $\lambda$, $\rho$) in (\ref{eq:lmm-knowledge}) also affect the performance of our model. In our experiments, we tune the parameters one by one using a validation set. During the tuning process, we observe that the performances are not sensitive to the values of $\lambda$ and $\rho$, but are sensitive to the values of $\alpha$ and $\beta$, which control the weights of semantic knowledge.

\subsection{Discussion}
We discuss some issues with regard to training of latent matching models.

\subsubsection{Parallelization}

\begin{figure}
  \centering
  \includegraphics[scale=0.5]{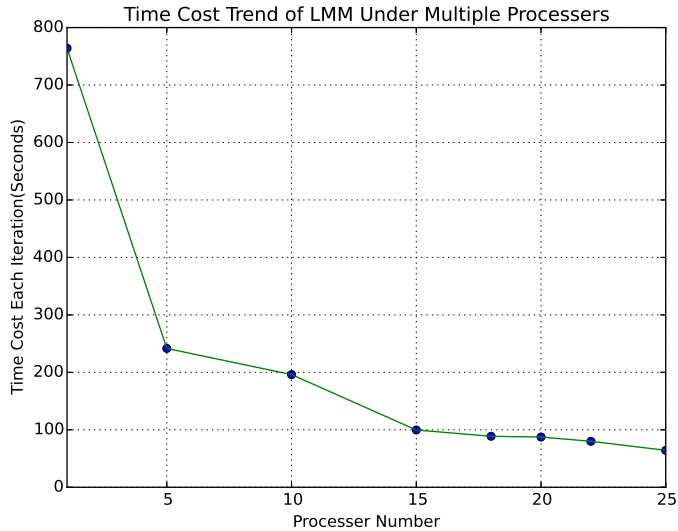}\\
  \caption{Time cost of each iteration in LMM under multiple processors (dimension=100)}\label{fig:processes}
\end{figure}

In our experiment, we mainly employ the coordinate descent algorithm in Algorithm \ref{alg:lmm-cd} and Algorithm \ref{alg:lmm-knowledge-cd}.
The matrix manipulation in each iteration can be computed in parallel, or more specifically, by using multi-core processors. Figure \ref{fig:processes} shows the time cost in each iteration versus the number of processors. We can clearly see that the training time can be substantially reduced with more processors used.

\subsubsection{Initialization}
The initialization of the mapping matrices $L_x$ and $L_y$ can affect the convergence rate of the learning algorithm. With random initialization, it usually takes 100 iterations to train an LMM. For training LMM augmented with semantic knowledge, we can directly utilize the result learned by LMM as the initial matrices. Experiment shows that much less iterations are needed to converge, and usually the number is less than 20 iterations.

\subsubsection{Coordinate Descent vs Gradient Descent}
In Section \ref{sec:lmm-ir}, we propose two kinds of optimization algorithms: coordinate descent and gradient descend. For coordinate descent, the convergence rate is faster, but in each step the computation of inverting two $d \times d$ matrices is expensive, especially when $d$ is large. For gradient descent, the computation at each step is faster, but it requires more steps to converge. Therefore, one needs to select a more suitable algorithm from the two in a specific setting. In our experiment setting, we find that coordinate descent is usually more efficient, and thus we use it as the main optimization method.

\subsubsection{Empirical Analysis}
We make analysis of the ranking results of LMM and LMM-X.  In many cases, we find that the semantic relations embedded in LMM-X can indeed improve relevance ranking. For example, the terms ``best'', ``most'', ``hardest'', and ``strongest'' are mined as synonyms from the log, and these terms are clustered together in the latent space induced by LMM-Synonyms.  In search, for the query of ``best game in the history'',  documents about ``most popular game'', ``hardest game'' and ``strongest game'' are promoted to higher positions, which enhances the relevance as well as the richness of search result. However, there are also some bad cases, mainly due to noises in the synonym dictionary. For example, in one experiment our mining algorithm identifies ``google'' and ``baidu'' as synonyms. Then for the query of ``google map'', a document about ``baidu map'' is ranked higher than a document about ``google earth''. Therefore, improving the quality of the mined semantic knowledge is one issue which we need to address in the future.

\section{Conclusion}\label{sec:conclusion}

In this paper, we have studied the problem of latent semantic matching for search. We have proposed a linear latent semantic model that leverages not only click-through data, but also semantic knowledge such as synonym dictionary and category hierarchy. The semantic knowledge is incorporated into the model by imposing regularization based on synonym and/or category information. We employ two methods to acquire semantic knowledge. One is to mine synonyms from the click bipartite graph, and the other is to utilize categories of documents. We have developed a coordinate descent algorithm and a gradient descent algorithm to solve the learning problem. The algorithms can be employed depending on the settings. We have conducted experiments on two large-scale datasets from a mobile app search engine. The experimental results demonstrate that our model can make effective use of semantic knowledge, and significantly outperform existing matching models.

Semantic matching on long tail data is still quite challenging in search and many other fields. In the future, we plan to work on developing matching models and algorithms for handling tail queries, which includes: (1) incorporating more types of knowledge such as knowledge graph into the matching model, and (2) incorporating semantic knowledge into a non-linear model.

\bibliographystyle{abbrv}
\bibliography{RL2M}
\end{document}